\documentclass[journal, twoside]{IEEEtran}
\usepackage{mathrsfs}  

\IEEEoverridecommandlockouts                              
\usepackage{mathrsfs}
\usepackage{graphicx}
\usepackage{subfigure}
\usepackage{url}
\usepackage{wrapfig}
\usepackage{algorithm}
\usepackage{algorithmic}
\usepackage{amsmath}
\usepackage{array}
\usepackage{dsfont}
\usepackage{amsfonts}
\usepackage{color}
\usepackage{amssymb}
\usepackage{multicol}
\usepackage{cite}
\usepackage{mathrsfs}
\usepackage{txfonts}
\usepackage{amsmath,amssymb}
\usepackage{tabularx}
\usepackage{multirow}
\allowdisplaybreaks
\floatname{algorithm}{\emph{Algorithm}}

\newcommand{\cS}{\mathcal{S}}

\newcommand{\lfteqn}{\begin{eqnarray} \begin{array}{lllllll}}
\newcommand{\ndeqn}{\end{array} \nonumber \end{eqnarray}}
\newcommand{\Lfteqn}{\begin{eqnarray} \begin{array}{lllllll}}
\newcommand{\Ndeqn}{\end{array}  \end{eqnarray}}
\def\bull{\vrule height 1.1ex width 1.1ex depth -.0ex }

\newtheorem{theorem}{Theorem}
\newtheorem{proposition}{Proposition}

\newtheorem{definition}{Definition}

\newtheorem{lemma}{Lemma}

\newtheorem{example}{Example}

\title{\LARGE \bf
Supervisory Control of Discrete Event Systems for Small Language Under Cyber Attacks}

\author{Xiaojun Wang,~\IEEEmembership{Member,~IEEE,}  Shaolong Shu,~\IEEEmembership{Senior Member,~IEEE}, and Feng Lin,~\IEEEmembership{Fellow,~IEEE} 
\thanks{This work is supported in part by the Natural Science Foundation of China under Grants 62403321 and 62473289; and by  the National Science Foundation of USA under grant 2146615.}
\thanks{X. Wang is with the School of Optical-Electrical and Computer Engineering, University of Shanghai for Science and Technology, Shanghai 200093, China (e-mail: wangxiaojun@usst.edu.cn).}
\thanks{S. Shu is with the School of Electronics and Information Engineering, Tongji University, Shanghai, China (e-mail: shushaolong@tongji.edu.cn).}
\thanks{F. Lin is with the Department of Electrical and Computer Engineering, Wayne State University, Detroit, MI 48202, USA (e-mail: flin@wayne.edu).}\thanks{Corresponding authors:  Xiaojun Wang and  Shaolong Shu.}}

\begin{document}

\maketitle

\begin{abstract}
	
Cyber attacks are unavoidable in networked discrete event systems where the plant and the supervisor communicate with each other via networks. Because of the nondeterminism in observation and control caused by cyber attacks, the language generated by the supervised system becomes nondeterministic. 
	The small language is defined as the lower bound on all possible languages that can be generated by the supervised system, which is needed for a supervised system to perform some required tasks under cyber attacks. In this paper, we investigate supervisory control for the small language. 
	After introducing CA-S-controllability and CA-S-observability, we prove that the supervisory control problem of achieving a required small language is solvable if and only if the given language is CA-S-controllable and CA-S-observable. If the given language is not CA-S controllable and/or CA-S-observable, we derive conditions under which the infimal CA-S-controllable and CA-S-observable superlanguage exists and can be used to design a supervisor satisfying the given requirement.

\end{abstract}



\section{Introduction}

With the rapid development of computer, communication and control technologies, wired or wireless communication networks are used to exchange information between  controllers and plants, which have the advantages of high efficiency, low cost, and high transmission capability. 
Since information is sent through networks, it is unavoidable that the controlled system may suffer from cyber attacks. Therefore, control of systems under cyber attacks has become an important research direction in systems and control, for both continuous-variable systems and discrete event systems. 

In this paper, we investigate supervisory control of discrete event systems (DES), where the plant and the supervisor are communicated via networks. 
Specifically, the communication channel from the plant to the supervisor is called observation channel and the communication channel from the supervisor to the plant is called control channel. 
In observation channels, an attacker can delete, insert, and/or change some observable events, which may cause the supervisors to make incorrect decisions. In control channels, cyber attacks may alter disablement or enablement of some controllable events, which will change the behavior of the system.

Within the DES framework, cyber attacks are investigated extensively \cite{ji2019enforcing, thorsley2006intrusion, carvalho2018detection, lima2017security, you2021supervisory}. For example, \cite{thorsley2006intrusion}  shows that an intruder may
interfere with the feedback performance of the system, causing the controllers in a
supervisory control system to fail. 
To solve the problem, the authors propose  a framework for modeling supervisory control systems to estimate how much damage that cyber attacks might cause.
In \cite{carvalho2018detection}, four types of cyber attacks are discussed: sensor insertion attacks (event insertion attacks), sensor erasure attacks (event deletion attacks), actuator enablement attacks (event enablement attacks), and actuator disablement attacks (event disablement attacks).  
It is shown that a diagnoser can be constructed to detect attacks. Once an attack is detected, the supervisor will disable all controllable events.

The researchers categorize  cyber attacks into the following three types: sensor
	attacks, actuator attacks, and joint sensor-actuator attacks. 
	Sensor attacks in observation channels are considered in \cite{meira2021synthesis, yao2024sensor, zhang2021joint,Tong2022polynomial, meira2020synthesis, su2018supervisor, tai2021synthesis, tai2022synthesis}. Among them, \cite{yao2024sensor} investigates the problem of synthesizing active sensor attackers against initial-secret of supervisory control systems. 
	Based on an all attack structure which records state estimates for both the supervisor and the attacker, the authors present algorithms for synthesizing successful attack strategies.    
\cite{zhang2021joint} investigates the problem of state estimation under attacks which may erase some events that have occurred and/or insert some events that have not actually occurred. 
The authors solve the problem by constructing a joint estimator which contains all the possible attacks. 
In \cite{Tong2022polynomial},  the authors propose a new attack detection mechanism in which the supervisor
only needs to keep track of the last observable event received to solve the problem of
detecting stealthy sensor attackers in cyber-physical discrete event systems. The authors of papers \cite{tai2021synthesis} and \cite{tai2022synthesis} propose to model sensor attackers with the following steps.  First, sensor attack constraints
are modeled as a finite state automaton $AC$, which describes the attack capabilities.  It is required that the sensor attack action (insertion, deletion, and replacement) initiated by the sensor attacker is instantaneous. Second, the sensor attack over attack constraint is modeled as a finite state automaton $A$, which is the attack that they aim to synthesize. Third, a fixed unit time interval, i.e., one $tick$, is used to model the observation channel. 

Actuator attacks in control channels are investigated in \cite{ma2022resilient, he2024estimation, carvalho2016detection,li2020detection}.   Among them, \cite{ma2022resilient} proposes the resiliency automata, based on which the authors develop a polynomial method to design a resilient supervisor such that the plant under control does not reach any unsafe state if the number of actuator attacks is less  than the required safety level. 
In \cite{he2024estimation}, the problems of estimation and prevention of actuator attacks are studied. Based on the proposed notions of strong and weak  actuator enablement estimabilities, the authors design an estimator and a prevention module to mislead an intruder’s attack estimation by using the reverse sensor functions to modify sensor readings. 
\cite{carvalho2016detection} considers the problem of intrusion detection and prevention in supervisory control systems, where the attacker has the ability to enable vulnerable actuator events that are disabled by the supervisor. To solve the problem, the authors  present a mathematical model and propose a defense strategy.


Joint sensor-actuator attacks in both observation and control channels are investigated in \cite{meira2023dealing, zhang2022sensor, lin2019towards, lin2021synthesis, lin2020synthesis, lima2021security, lima2019security, hadjicostis2022cybersecurity}. 
In \cite{meira2023dealing}, the authors consider a supervisor that guarantees the safety of the system even when sensor readings and actuator commands are compromised. Their solution methodology reduces the problem of synthesizing a robust supervisor against deception attacks to a conventional supervisory control problem. 
In \cite{zhang2022sensor}, the authors develop an attack structure computed as the parallel composition of the attacker observer and the supervisor under attack. It is used  to select attacks that cause the closed loop system to reach an unsafe state. 
In \cite{lin2019towards}, the authors synthesize resilient supervisors against combined actuator and sensor attacks. A constraint-based approach for the bounded synthesis of resilient supervisors is developed by reducing the problem to a quantified Boolean formula problem.
In \cite{lin2021synthesis} and \cite{ lin2020synthesis},  the authors investigate how to find a powerful joint sensor and actuator attack policy, not how to synthesize a supervisor to tame attacks. For more information on cyber attacks in discrete event systems, the reader is referred to a tutorial paper [25].

We have also investigated supervisory control of DES under joint sensor-actuator attacks. For sensor attacks, we propose a new attack model, called ALTER (Attack Language for Transition-basEd Replacement) model \cite{zheng2021modeling, zheng2024modeling, lin2023diagnosability, lin2024diagnosability}. 
In the ALTER model, an attackable transition can be replaced by any strings in the corresponding attack language. The  ALTER model is  both general and specific. 
It is general in the sense that common sensor attacks such as deletions, insertions, replacements, and all-out attacks can all be modeled by the ALTER model. It is specific in the sense that all attacks are specified by the attack languages.

It is shown in \cite{zheng2024modeling} that, due to nondeterministic attacks, the language generated by the supervised system is nondeterministic.
Not knowing this language can cause serious problems in networked supervisory control. This is because, given a legal language, if we do not know the language
generated by the supervised system, then we do not know if the system is safe or not. Similarly, given a required language, if we do not
know the language generated by the supervised system, then we do not know if the system can perform some basic tasks described by
the required language or not. To handle this new situation, the upper bound (called large language) and the lower bound (called small language) on all possible languages generated by the supervised system are defined. 
The large language is needed to guarantee the safety of the supervised system. The safety problem using the large language is solved in \cite{zheng2024modeling} by extending controllability and observability to CA-controllability and CA-observability. 

The small language is needed to ensure that the supervised system can always perform some basic tasks, because if we know that the small language contains the required language describing the basic tasks, then we know that the system can
	perform these basic tasks, no matter which language it actually generates. The small language has not been investigated until this paper. It plays an important role in ensuring that some required tasks described by a required language $K_r$ can be performed by the supervised system under all possible cyber attacks. 
Note that this cannot be done using the large language, because even if the required  language  $K_r$  is contained in the large language, there is still no guarantee that $K_r$ is contained the actual language generated by the supervised system, as the large language is the upper bound, not the lower bound.

In this paper, we investigate the small language under joint sensor-actuator attacks. Our approach is as follows. First, we introduce two new concepts, called CA-S-controllability and CA-S-observability, to  obtain necessary and sufficient conditions for the existence of a supervisor whose small language is equal to a given specification language, under joint sensor-actuator attacks. Second, we  investigate how to synthesize a supervisor such that some required tasks specified by a required language $K_r$ can always be performed by the supervised system, even under cyber attacks when $K_r$ is CA-S-controllable and CA-S-observable. Third, we prove that both CA-S-controllability and CA-S-observability are preserved under language intersection. Fourth,  if $K_r$ is not   CA-S-controllable and$/$or CA-S-observable, we find, if possible, the infimal CA-S-controllable and CA-S-observable superlanguage of $K_r$ and synthesize a supervisor whose small language is equal to the superlanguage.

Unfortunately, unlike in conventional supervisory control, the infimal CA-S-controllable and CA-S-observable superlanguage of $K_r$ does not always exist. 
This is because the plant language $L(G)$ itself may not be CA-S-controllable, which is contrary to conventional supervisory control, where $L(G)$ is always controllable and observable. This counter-intuitive result makes the small language much more difficult to handle. 
To overcome this difficulty, we calculate the largest sublanguage  of $L(G)$ that is CA-S-controllable and denote it by $L_{na}(G)$. We show that if $K_r \subseteq L_{na}(G)$, then the infimal CA-S-controllable and CA-S-observable superlanguage of $K_r$ always exists. This new approach has never been used in supervisory control before.


The paper is structured as follows.  Section II introduces DES and cyber attacks.  Section III formally states the supervisory control problem of DES to achieve a required language under cyber attacks.  
Section IV sloves the supervisory control problem under cyber attacks. Section V investigates  the infmal  CA-S-controllable and CA-S-observable superlanguage. Some concluding remarks are given in Section VI.



\section{Discrete Event Systems under Cyber Attacks}

In this section, we briefly review the results on DES and cyber attacks introduced in \cite{wonham2019supervisory, cassandras2009introduction,  zheng2024modeling}. 

\subsection{Discrete event systems}

A DES is modeled by a finite deterministic automaton 
$$
G=( Q, \Sigma, \delta, q_0, Q_m),
$$
where $Q$ is the set of states; $\Sigma$ is the set of events; $\delta : Q \times \Sigma \rightarrow Q$ is the (partial) transition function; $q_0$ is the initial state; and $Q_m$ $\subseteq$ $Q$ is the set of  marked states. 
The set of all possible transitions is also denoted by $\delta$: $\delta = \{(q, \sigma, q'): \delta (q, \sigma) = q' \}$. Denote  the set of all strings over $\Sigma$ by $\Sigma ^*$. 
The language {\em generated} by $G$ is the set of all strings defined in $G$ from the initial state, that is,
$$L(G) = \{ s \in \Sigma ^*: \delta (q_0,s)! \},$$
where ``!'' means ``is defined''.  The language {\em marked} by $G$ is defined as
$$L_m(G)= \{s\in L(G): \delta(q_0,s) \in Q_m\}.$$

In general, a language $K \subseteq \Sigma ^*$ is a set of strings. For a string $s \in \Sigma ^*$, we use $s' \leq s$ to denote that $s'$ is a prefix of $s$. 
The length of $s$ is denoted by $|s|$.  The (prefix) closure of $K$, denoted by $\overline{K}$, is the set of all prefixes of strings in $K$. A language is (prefix) closed if it equals its prefix closure. By the definition, $L(G)$ is closed. 

A controller, called supervisor, is used to control the system, called plant, so that some objective is achieved. The supervisor can control some events and observe some other events.  
The set of controllable events is denoted by $\Sigma _c$ ($\subseteq \Sigma$), $\Sigma _{uc} = \Sigma - \Sigma _c$ is the set of uncontrollable events. 
The set of observable events is denoted by $\Sigma _o$ ($\subseteq \Sigma$). $\Sigma _{uo} = \Sigma - \Sigma_o$ is the set of unobservable events. 
The set of observable transitions is denoted by $\delta_o$: $\delta_o = \{ (q, \sigma, q'): \delta (q, \sigma) =q' \wedge \sigma \in \Sigma _o\}$; and the set of unobservable transitions is denoted by $\delta_{uo}$: $\delta_{uo} = \{ (q, \sigma, q'): \delta (q, \sigma) =q' \wedge \sigma \in \Sigma _{uo}\}$.  

For a given string, its observation is described by the natural projection  $P: \Sigma^* \rightarrow \Sigma^*_o$, which is  defined as
\lfteqn
P(\varepsilon) = \varepsilon\\
P(\sigma)=\left\{ \begin{array}{ll}
	\sigma & \mbox{ if }\sigma \in \Sigma_o\\
	\varepsilon   & \mbox{ if } \sigma \in \Sigma_{uo} \\
\end{array}\right.\\ 
P(s\sigma)= P(s)P(\sigma), s\in \Sigma^*,\sigma \in \Sigma,
\ndeqn
where $\varepsilon$ is the empty string.

\subsection{Cyber Attacks in Observation Channel}

Cyber attacks are unavoidable in communication channels. We use the ALTER attack model proposed in \cite{zheng2021modeling, zheng2024modeling, lin2023diagnosability, lin2024diagnosability} to describe sensor attacks in the observation channel 
as follows\footnote{How to implement the ALTER model using automata is discussed in \cite{zheng2024modeling, lin2024diagnosability}. The reader is refereed to \cite{zheng2024modeling, lin2024diagnosability} for more details. }.
The set of observable events and transitions that can be attacked, called attackable events and attackable transitions, are denoted by $\Sigma^a_o$ $\subseteq$ $\Sigma_o$ and $\delta^a$ $=$ $\{(q,\sigma, q')\in \delta :\sigma\in \Sigma_o^a\}$, respectively. 

For an attackable transition $tr$ $=$ $(q, \sigma, q')$ $\in$ $\delta^a$, we assume that an attacker can change the event $\sigma$ to any string in the corresponding attack language $A_{tr}$ $\subseteq$ $\Sigma_o^*$. 
$A_{tr}$ can be determined based on the information of the attacker. In particular, the ALTER model can handle deletion (by letting $A_{(q, \sigma, q')} = \{ \varepsilon, ... \}$), replacement (by letting $A_{(q, \sigma, q')} = \{ \alpha, ... \}$), insertion (by letting $A_{(q, \sigma, q')} = \{ \alpha \sigma, \sigma \alpha, ... \}$), all-out attacks (by letting $A_{(q, \sigma, q')} = \Sigma _o^*$), and so on. There may be  more than one attackable transitions, denote the set of all attack language as $$\mathbb{A} = \{A_{tr}: tr= (q, \sigma, q') \in \delta^a\}.$$
Note that $\mathbb{A}$ contains all attack languages. Each attack language may contain more than one strings, which makes the attacks nondeterministic. Cyber attacks can then be modeled by a mapping from the set of attackable transitions to the set of attack languages as
$$
\pi : \delta ^a\rightarrow \mathbb{A},
$$
where $\pi(tr)=A_{tr}$.  

If a string $s = \sigma_1\sigma_2\cdots\sigma_{|s|}$ $\in$ $L(G)$ occurs in $G$, the set of all possible strings after  cyber attacks, denoted by $\Theta ^\pi(s)$, is obtained as follows. Denote $q_k$ $=$ $\delta(q_0, \sigma_1\cdots\sigma_k)$, $k$ $=$ $1,2,\cdots, |s|$, then
$$
\Theta ^\pi(s) = L_1 L_2 ... L_{|s|},
$$
where
\begin{equation} \nonumber
	\begin{split}
		L_k = \left\{ \begin{array}{ll}
			\{ \sigma _k \} & \mbox{if } (q_{k-1}, \sigma _k, q_k) \not\in \delta ^a \\
			A_{(q_{k-1}, \sigma _k, q_k)} & \mbox{if } (q_{k-1}, \sigma _k,
			q_k) \in \delta ^a. \end{array} \right .
	\end{split}
\end{equation}
Note that $\Theta ^\pi(s)$ may contain more than one string. Hence,   $\Theta ^\pi$  is a mapping:
\Lfteqn
\Theta ^\pi  : L(G) \rightarrow 2^{\Sigma^*}. \nonumber
\Ndeqn

The observation under both partial observation and cyber attacks in the observation channel is then given by 
$$
\Phi ^\pi = P \circ \Theta ^\pi,
$$
where $\circ$ denotes composition of functions. In other words, for $s \in L(G)$, $\Phi ^\pi (s) = P ( \Theta ^\pi (s))$. Hence, $\Phi ^\pi$ is a mapping from $L(G)$ to $2^{\Sigma _o^*}$:
$$
\Phi ^\pi: L(G) \rightarrow 2^{\Sigma _o ^*}.
$$

We extend $P$, $\Theta ^\pi$, and $\Phi ^\pi$ from strings $s$ to languages $L$ in the usual way as
\begin{equation} \nonumber
	\begin{split}
		& P(L) = \{ t \in \Sigma^*_o: (\exists s \in L) t = P(s) \} \\
		& \Theta ^\pi(L) = \{ t \in \Sigma^*: (\exists s \in L) t  \in
		\Theta ^\pi(s) \} \\
		& \Phi ^\pi(L) = \{ t \in \Sigma^*_o: (\exists s \in L) t  \in \Phi
		^\pi(s) \}.
	\end{split}
\end{equation}

After the occurrence of $s \in L(G)$, the string observed by the supervisor $\cS$ is one of the strings in $ \Phi ^\pi (s)$, denoted as $t \in \Phi ^\pi (s)$. 
The state estimate after observing $  t \in \Phi^\pi (s)$ is denoted as $SE^\pi_G(t)$, which is defined as
\begin{equation} \nonumber
	\begin{split}
		SE^\pi_G(t)= \{& q \in Q: (\exists s \in L(G)) \\
		& t \in \Phi ^\pi (s) \wedge \delta  (q_0, s) = q \}.
	\end{split}
\end{equation}

Let us recall the steps in \cite{zheng2024modeling} to obtain the state estimates.

{\em Step 1:}  For each attackable transition $tr\in \delta^a$,  let $A_{tr} = L_m (F_{tr})$ for some
$$
F_{tr} = (Q_{tr} , \Sigma, \delta_{tr}, q_{0,tr}, Q_{m,tr}).
$$

{\em Step 2:}  Replace an attackable transition $tr=(q, \sigma, q') \in \delta ^a$ in $G$ by $(q, F_{tr}, q')$ as follows.
\begin{align*}
	G_{tr \rightarrow (q, F_{tr}, q')} = ( Q \cup
	Q_{tr}, \Sigma , \delta_{tr \rightarrow (q,
		F_{tr}, q')}, q_0 ),
\end{align*}
where $\delta_{tr \rightarrow (q, F_{tr}, q')} = (\delta - \{ (q, \sigma, q') \} ) \cup \delta_{tr} \cup \{ (q, \varepsilon, q_{0,tr}) \} \cup \{ (q_{m,tr}, \varepsilon, q'): q_{m,tr} \in Q_{m,tr} \}$.

Denote the automaton after replacing all attackable transitions as
$$
G^\diamond=( Q^\diamond, \Sigma , \delta^ \diamond , q_0, Q^\diamond_m)=( Q \cup \hat{Q}, \Sigma , \delta^ \diamond , q_0, Q),
$$
where $\hat{Q}$ is the set of states added during the replacement and $Q^\diamond_m=Q$ is the set of marked states. 
Note that $G^\diamond$ is a nondeterministic automaton, that is, $\delta^ \diamond$ is a mapping $\delta^ \diamond : Q^\diamond \times \Sigma \rightarrow 2^{Q^\diamond}$.

{\em Step 3:}   Replace unobservable
transitions in $G^\diamond$ by $\varepsilon$-transitions and denote the resulting automaton as
$$
G_{\varepsilon}^\diamond=( Q \cup \hat{Q}, \Sigma_o , \delta _{\varepsilon}^\diamond , q_0, Q),
$$
where $\delta _{\varepsilon}^\diamond = \{ (q, \sigma, q'): (q, \sigma, q') \in \delta ^\diamond \wedge \sigma \in \Sigma _o \} \cup \{ (q, \varepsilon, q'): (q, \sigma, q') \in \delta ^\diamond \wedge \sigma \not\in \Sigma _o \} $.  

{\em Step 4:} Convert $G_{\varepsilon}^\diamond$ to CA-observer $G_{obs}^\diamond$ using operator $OBS$ as follows.
\begin{equation}
	\begin{aligned}
		G_{obs}^\diamond & = OBS(G_{\varepsilon}^\diamond) =(X,\Sigma _o, \xi, x_0, X_m) \\
		&= Ac(2^{Q \cup \hat{Q}},\Sigma _o,
		\xi, UR(\{ q_0\}), X_m), \label{OBS}  \notag
	\end{aligned}
\end{equation}
where $Ac(\cdot)$ denotes the accessible part; $UR(\cdot)$ is the unobservable reach defined, for $x \subseteq Q \cup \hat{Q}$, as
$$
UR(x) = \{ q \in Q \cup \hat{Q}: (\exists q' \in x) q\in \delta
_{\varepsilon}^\diamond (q', \varepsilon) \}.
$$

The transition function $\xi$ is defined, for $x \in X$ and $\sigma \in \Sigma _o$ as
$$
\xi (x, \sigma) = UR(\{q \in Q \cup \hat{Q}: (\exists q' \in
x)q\in \delta _{\varepsilon}^\diamond (q',\sigma) \}).
$$
The marked states are defined as
$$
X_m = \{x \in X: x \cap Q \not = \emptyset \}.
$$

It is shown in \cite{zheng2024modeling} that
$$
\Phi ^\pi (L(G))=L_m(G_{obs}^\diamond).
$$

The following theorem is proved in \cite{zheng2024modeling}.
\begin{theorem} \rm
	\label{theorem1} 
Consider a discrete event system $G$ under cyber attacks. After observing $t \in \Phi ^\pi (L(G))
	=L_m(G_{obs}^\diamond)$, the state estimate $SE^\pi_G(t)$ is given by 
	\begin{align}
		SE^\pi_G(t)= \xi (x_0, t) \cap Q. 
	\end{align}
\end{theorem}

\subsection{Cyber Attacks in Control Channel}

We assume that the disablement$/$enablement status of some controllable events can be changed by an attacker in the control channel. In other words, an attacker can enable an event that is disabled by the supervisor and/or disable an event that is enabled by the supervisor. 
Denote the set of attackable controllable events by $\Sigma^a_c$ $\subseteq$ $\Sigma_c$. Note that uncontrollable events are always permitted to occur and no attacker can disable them.

Based on its observation $t \in \Phi ^\pi (s)$, supervisor $\cS$ enables a set of events, denoted by $\cS(t)$. Hence, $\cS$ is a mapping, 
$$
\cS: \Phi ^\pi(L(G)) \rightarrow 2 ^\Sigma.
$$
Note that we require $\Sigma_{uc}$ $\subseteq$ $\cS(t)$ because uncontrollable events cannot be disabled.

Under cyber attacks in control channel, for a given control $\gamma \in 2^\Sigma$, some events in $\Sigma^a_c$ can be added to it or removed from it. Hence, the possible controls are:
\lfteqn 
\Delta (\gamma) =\{\gamma_a \in 2^{\Sigma} : (\exists \gamma', \gamma'' \subseteq \Sigma^a_c) \ \gamma_a = (\gamma-\gamma')\cup \gamma''\}.
\ndeqn

When the supervisor issues a control command $\cS(t)$ after observing $t \in \Phi ^\pi(L(G))$, it may be altered under cyber attacks. 
We use $\cS^a(t)$ to denote the set of all possible control commands  that may be received by the plant under cyber attacks, that is,
\lfteqn 
\cS^a(t) = \Delta(\cS(t)).
\ndeqn

Let us illustrate the results under cyber attacks using the following example.

\begin{example}  \rm
	
	Consider the discrete event system $G$ shown in Fig.~\ref{fig1}. Assume that $\eta$ is unobservable and all events are controllable, that is, $\Sigma_o = \{\alpha, \beta, \mu,\lambda\}$ and  $\Sigma_c=\Sigma$. 
	Observations of events $\alpha$ can be changed by an attacker, that is, $\Sigma^a_o = \{\alpha\}$. 
	The attack language for transition $tr$ $=$ $(3,\alpha, 4)$ is $A_{tr}$ $=$ $\{\varepsilon, \alpha, \alpha\alpha\}$, where $\varepsilon$ (resp., $\alpha\alpha$) corresponds to deletion attack (resp., insertion attack). The attacker can also enable$/$disable the occurrence of events  $ \beta$ and $\lambda$, that is, $\Sigma^a_c$ $=$ $\{ \beta,\lambda\}$.  
	
	\begin{figure}[hbt!]
		\centering
		\includegraphics[height=0.8 in]{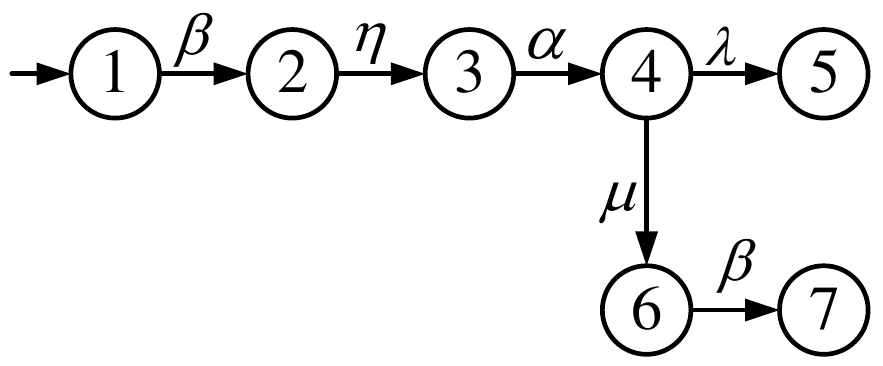}
		\caption{A discrete event system $G$.}
		\label{fig1}
	\end{figure}	
	 
	Let us consider the possible observations for the string $s= \beta\eta\alpha$. By the definition, we have
	$$ 
	\Theta ^\pi (s)= \{ \beta\eta\}A_{tr} = \{ \beta\eta,  \beta\eta\alpha,  \beta\eta\alpha\alpha\}.
	$$
	
	Since $\eta$ is unobservable, we then have 
	$$
	\Phi ^\pi(s) = P(\Theta ^\pi (s))  = \{ \beta, \beta\alpha, \beta\alpha\alpha\}.
	$$
	
	If $t= \beta\alpha$ is observed, the supervisor issues a control command $\cS( \beta\alpha) = \{ \mu,\lambda\}$, However, the actual control command received by the plant is one of the following
	$$
	\cS^{a}(t)=\{\{\mu\}, \{ \mu,\lambda\}, \{ \mu, \beta\}, \{\mu, \lambda,  \beta\}\}.
	$$
	
	Without cyber attacks, $\lambda$ is enabled by the supervisor after observing $t= \beta\alpha$. However, if the control command received by the plant is $\{\mu\}$, then $\lambda$ is disabled by the attacker.
				
\end{example}

\section{Problem Statement}

\begin{figure}[hbt!]
	\centering
	\includegraphics[height=1.8 in]{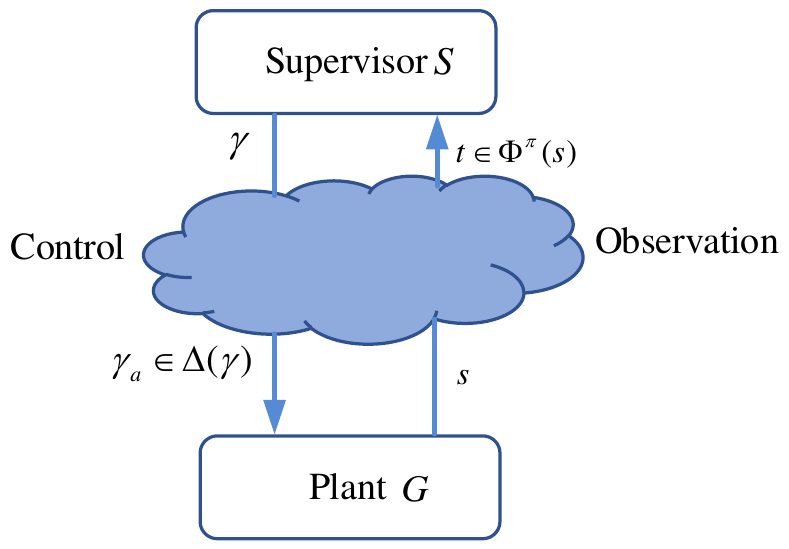}
	\caption{The structure of supervisory control under cyber attacks.}
	\label{fig2}
\end{figure}

As discussed in the previous section, cyber attacks can happen in both observation and control channels as shown in Fig.~\ref{fig2}. 
When a string $s \in L(G)$ occurs in the plant, an attacker can change the observation of string $s$ from $P(s)$ to one of the string in $\Phi ^\pi(s)$, that is, $t \in \Phi ^\pi(s)$. 
Based on the observation $t$, $\cS$ issues a control command $\cS(t)$, which may be altered to any control command in $\cS^a(t)$ by the attacker. 

The supervised system under cyber attack is denoted as $\cS^{a} / G$. The language generated by the supervised system, denoted by  $L(\cS^a/G)$, is nondeterministic. 
The reasons for nondeterminism are as follows: (1) for a given string $s \in L(G)$ occurred in $G$, the string $t \in \Phi ^\pi(s)$ observed by $\cS$ is nondeterministic and 
(2) the  control command $\cS^a(t)$ being used by $G$ is also nondeterministic. Hence, the language is not unique. Of all possible languages $L(\cS^a/G)$, their lower bound is called small language and denoted as $L_r(\cS^a/G)$
\cite{zheng2024modeling}. $L_r (\cS^a/G)$ is defined recursively as follows.

\begin{enumerate}
	\item
	The empty string belongs to $L_r (\cS^a/G)$:
	$$
	\varepsilon \in L_r (\cS^a/G).
	$$
	\item
	If $s$ belongs to $L_r (\cS^a/G)$, then for any $\sigma \in \Sigma$, $s \sigma$ belongs to $L_r (\cS^a/G)$ if and only if $s \sigma$ is allowed in $L(G)$ and $\sigma$ is uncontrollable or enabled by $\cS^a$ in {\em all} situations:
	\begin{align*}
		& (\forall s \in L_r (\cS^a/G))(\forall \sigma \in \Sigma) s \sigma \in L_r (\cS^a/G) \\
		\Leftrightarrow 
		& s \sigma \in L(G) \wedge (\sigma \in \Sigma _{uc} \vee (\forall t \in   \Phi ^\pi(s)) \\
		& (\forall \gamma_a \in  \cS^a(t))\sigma \in \gamma_a).
	\end{align*}	
\end{enumerate}

The upper bound on all possible languages is called large language and denoted as  $L_a (\cS^a/G)$\cite{ zheng2024modeling}.  $L_a (\cS^a/G)$ is defined recursively as follows.

\begin{enumerate}
	\item
	The empty string belongs to $L_a (\cS^a/G)$:
	$$
	\varepsilon \in L_a (\cS^a/G).
	$$
	\item
	If $s$ belongs to $L_a (\cS^a/G)$, then for any $\sigma \in \Sigma$,
	$s \sigma$ belongs to $L_a (\cS^a/G)$ if and only if $s \sigma$ is
	allowed in $L(G)$ and $\sigma$ is uncontrollable or enabled by
	$\cS^a$ in {\em some} situations:
	\begin{align*}
		& (\forall s \in L_a (\cS^a/G))(\forall \sigma \in \Sigma) s \sigma \in L_a (\cS^a/G) \\
		\Leftrightarrow & s \sigma \in L(G) \wedge (\sigma \in \Sigma _{uc} \vee (\exists t \in \Phi ^\pi(s)) \\
		& (\exists \gamma_a \in  \cS^a(t))\sigma \in \gamma_a) .
	\end{align*}
	
\end{enumerate}

While the large language is used to ensure that the supervised system never generate illegal string and/or enter unsafe states \cite{zheng2024modeling}, the small language is used to ensure that the supervised system can always perform some (minimally) required tasks described by a required language $K_r \subseteq L(G)$.

We investigate small language in this paper, that is, the goal of supervisory control is to ensure that the supervised system can always generate all strings in $K_r$, even under cyber attacks. 
Formally, this means that we would like to design a supervisor $\cS$, if possible, such that $K_r \subseteq L_r(\cS^a/G)$.


To this end, we first investigate the existence condition of a supervisor $\cS$ such that $L_r(\cS^a/G)=K$, where $K \subseteq L(G)$ is a given specification language.
Without loss of generality, we assume that $K$ is generated by a sub-automaton $H \sqsubseteq G$, that is, $K = L(H)$ for some 
$$
H=( Q_H, \Sigma , \delta_H , q_0 ),
$$
where $Q_H \subseteq Q$ and $\delta_H = \delta |_{Q_H \times \Sigma} \subseteq \delta$. Thus, the state set $Q$ is partitioned into required states  $Q_H$ and the rest of states  $Q-Q_H$.

We construct the CA-observer for $H$ in the same way as that for $G$ and denote the results by
\begin{align*}
	& H^\diamond=( Q_H^\diamond, \Sigma , \delta^\diamond_H, q_0, Q^\diamond_{Hm})=( Q_H \cup \hat{Q}, \Sigma, \delta_H^ \diamond, q_0, Q_H).\\
	& H_{obs}^\diamond = OBS(H_{\varepsilon}^\diamond) =(X_H,\Sigma_o, \xi_H, x_0, X_{Hm}).
\end{align*} 

In the rest of the paper, we will use subscript $H$ to denote things related to $H$. For example, $\delta_H^ \diamond$ denotes the transition function for $H^\diamond$.

Let us use the the following example to illustrate the necessity of a new approach to supervisory control under cyber attacks.

\begin{example} 
	
Again consider the DES $G$ shown in Fig.~\ref{fig1}. Assume that $\Sigma_o = \{\alpha, \beta, \mu,\lambda\}$, $\Sigma_c=\Sigma$, $\Sigma^a_o = \{\alpha\}$, and $\Sigma^a_c$ $=$ $\{ \mu\}$.  Let $K=L(H)$, where $H$ is the  sub-automaton of $G$ shown in Fig.~\ref{fig3}.

\begin{figure}[hbt!]
	\centering
	\includegraphics[height=0.8 in]{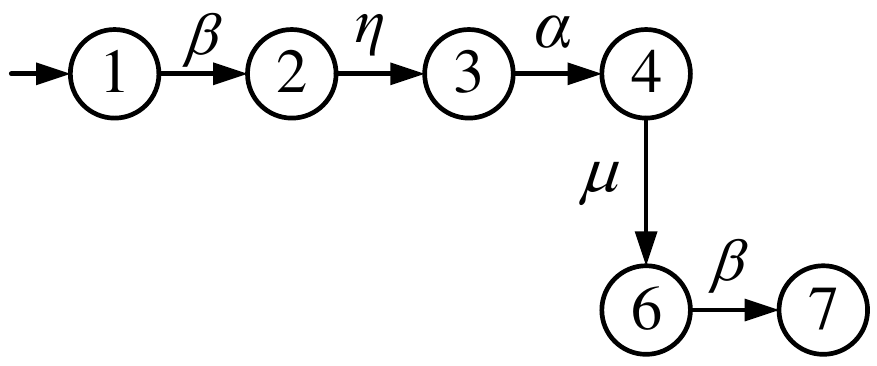}
	\caption{Subautomaton $H$ of $G$ with $L(H) = K$.}
	\label{fig3}
\end{figure}

Without cyber attacks, for string $s$ $=$ $\beta\eta\alpha$, $P(s)$ $=$ $\beta\alpha$,  a conventional supervisor will disables $\lambda$ and enables $\mu$ to achieve $K$, that is, $L(\cS/G) = K$. 
When there are cyber attacks in the system, $\mu$  can be disabled by an attacker. This leads to $L_r(\cS^a/G) = \overline{\beta \eta \alpha} \neq K$, which violates the specification.
Hence, the conventional method for supervisory control does not work under cyber attacks. A new method is needed for supervisory control under cyber attacks.


\end{example}  

\vspace{0.1in}

Formally, let us solve the following supervisory control problem.

\textbf{Supervisory Control Problem of Discrete Event Systems to Achieve a Required Language under Cyber Attacks  (SCPDES-RL-CA):} 
Consider a discrete event system $G$ under cyber attacks in the observation channel described by $\Phi^\pi$, and in the control channel described by $\Delta$. 
For a non-empty closed specification language $K \subseteq L(G)$ generated by a sub-automaton $H \sqsubseteq G$, find a supervisor $\cS: \Phi ^\pi(L(G)) \rightarrow 2 ^\Sigma$ such that $L_r(\cS^a/G) = K$.

\section{Problem Solutions}

Now, let us investigate how to solve SCPDES-RL-CA. We recall controllability \cite{ramadge1987supervisory} and observability \cite{lin1988observability} as follows. 

A closed language $K \subseteq L(G)$ is {\em controllable} with respect to $L(G)$ and $\Sigma _c$ if   
\begin{align*}
	K \Sigma _{uc} \cap L(G) \subseteq K.
\end{align*}

 A closed language $K \subseteq L(G)$ is {\em observable} with respect to $L(G)$ and $\Sigma _o$ if 
 \begin{align*}
	& (\forall s, s' \in K)(\forall \sigma \in \Sigma)(P(s) = P(s') \\
	&\wedge s\sigma \in L(G) \wedge s'\sigma \in K )\Rightarrow s\sigma \in K.
\end{align*}

We extend controllability to CA-S-controllability for supervisory control under cyber attacks for small language as follows. 
\begin{definition}
A closed nonempty language $K \subseteq L(G)$ is  CA-S-$controllable$ with respect to $L(G)$, $\Sigma_{uc}$, and $\Sigma^a_c$ if
\begin{align}
	\label{Defc}
	K \Sigma_{uc} \cap L(G) \subseteq K \wedge K \subseteq (\Sigma - \Sigma^a_c)^*.
\end{align}
\end{definition}

We extend observability to CA-S-observability for supervisory control under cyber attacks for small language as follows. 
\begin{definition}
A closed nonempty language $K \subseteq L(G)$ is  CA-S-$observable$ with respect to $L(G)$, $\Sigma_o$, $\Sigma^a_o$, and  $\Phi ^\pi$ if
\begin{align}
	\label{NSO}
	& (\forall s \in K)(\forall \sigma \in \Sigma) ((s\sigma \in L(G)  \nonumber\\
	& \wedge(\forall t \in \Phi ^\pi(s)) (\exists s' \in K) \nonumber\\
	&  t \in  \Phi ^\pi(s') \wedge s'\sigma \in K) \Rightarrow s\sigma \in K). 
\end{align} 
\end{definition}

Clearly, if there are no cyber attacks in the control channel, that is, $\Sigma _c^a = \emptyset$, then CA-S-controllability
reduces to controllability. Furthermore, we can prove the following proposition.

\begin{proposition} 
	\label{rem1} 

	If there are no cyber attacks in the observation channel, that is, $\Phi ^\pi(s) = P(s)$,  then  CA-S-observability reduces to  observability. 
	
\end{proposition}
\noindent {\em Proof:} 

Assume $\Phi ^\pi(s) = P(s)$. Then
\begin{align*}
	& K \mbox{ is CA-S-observable} \\
	\Leftrightarrow 
	&(\forall s \in K)(\forall \sigma \in \Sigma) ((s\sigma \in L(G) \\
	& \wedge(\forall t \in \Phi ^\pi(s)) (\exists s' \in K) \\
	&  t \in \Phi ^\pi(s') \wedge s'\sigma \in K) \Rightarrow s\sigma \in K)\\
	\Rightarrow 
	& (\forall s \in K)(\forall \sigma \in \Sigma) ((s\sigma \in L(G) \wedge (\exists s' \in K) \\
	& P(s) = P(s') \wedge s'\sigma \in K) \Rightarrow s\sigma \in K) \\
	&(\mbox{because }\Phi ^\pi(s) =\{ P(s) \}\mbox{ is unique})\\
	\Leftrightarrow 
	& (\forall s \in K)(\forall \sigma \in \Sigma)(((\exists s' \in K) P(s) = P(s') \\
	&\wedge s\sigma \in L(G) \wedge s'\sigma \in K)\Rightarrow s\sigma \in K)\\
	\Leftrightarrow 
	&(\forall s \in K)(\forall \sigma \in \Sigma) (\neg ((\exists s' \in K) P(s) = P(s') \\
	&\wedge s\sigma \in L(G) \wedge s'\sigma \in K) \vee s\sigma \in K) \\
	\Leftrightarrow 
	&(\forall s \in K)(\forall \sigma \in \Sigma) ((\forall s' \in K) \neg(P(s) = P(s') \\
	&\wedge s\sigma \in L(G) \wedge s'\sigma \in K) \vee s\sigma \in K)\\
	\Leftrightarrow 
	& (\forall s \in K)(\forall \sigma \in \Sigma)((\forall s' \in K) ((P(s) = P(s') \\
	&\wedge s\sigma \in L(G) \wedge s'\sigma \in K) \Rightarrow s\sigma \in K)) \\
	\Leftrightarrow 
	& (\forall s, s' \in K)(\forall \sigma \in \Sigma)((P(s) = P(s') \\
	&\wedge s\sigma \in L(G) \wedge s'\sigma \in K )\Rightarrow s\sigma \in K) \\
	\Leftrightarrow
	& K \mbox{ is observable} .
\end{align*}
\hfill \bull

\begin{lemma}
	\label{lemmao}
	\rm
	If $K$ is  CA-S-observable with respect to $L(G)$, $\Sigma_o$, $\Sigma^a_o$, and  $\Phi ^\pi$, then, for all $s \in K$ and $\sigma \in \Sigma$,
	\begin{equation} \nonumber
		\begin{split}
			& (s\in K \wedge s \sigma \in L(G) \wedge (\forall t \in  \Phi ^\pi(s)) (\exists s' \in K) \\
			& t \in \Phi ^\pi(s') \wedge s' \sigma\in K) 
			\Leftrightarrow s \sigma \in K.
		\end{split}
	\end{equation}
\end{lemma}
\noindent {\em Proof:} 

($\Rightarrow$) This implication holds because of the definition of  CA-S-observable.

($\Leftarrow$) This implication can be proved as follows.
\begin{align*}
	& s \sigma \in K \\
	\Rightarrow
	& s\in K \wedge s \sigma \in L(G) \wedge s \sigma \in K \\
	\Rightarrow
	& s\in K \wedge s \sigma \in L(G) \wedge (\forall t \in  \Phi ^\pi(s)) (\exists s' \in K) \\
	& t \in \Phi ^\pi(s') \wedge s' \sigma\in K \\
	& (\mbox{let } s'=s).
\end{align*} 
\hfill \bull

Let us construct a state-estimate-based supervisor $\cS_p$, which is defined as
\begin{align}
	\label{MB}
	\cS_p (t) = \{\sigma \in \Sigma : (\exists q \in
	SE^\pi_H(t))  \delta_{H} (q,\sigma) \in Q_H \},
\end{align}
where 
\begin{align}
		\label{SE}
{SE}^\pi_H(t) = \{ q \in Q_H: ( \exists s \in L(H)) t \in \Phi ^\pi (s) \wedge \delta_H (q_0, s) = q\}
\end{align}
can be calculated using $H_{obs}^\diamond$ as 
$$
{SE}^\pi_H(t) = \xi_H (x_0, t) \cap Q_H.
$$

We then have the following theorem for SCPDES-RL-CA.
\begin{theorem} 
	\label{theo2}
	\rm
	Consider a discrete event system $G$ under cyber attacks. For a nonempty closed language $K \subseteq L(G)$, SCPDES-RL-CA is solvable if and only if (1) $K$ is CA-S-controllable with respect to $L(G)$, $\Sigma _{uc}$, and $\Sigma^a _{c}$; 
	and (2) $K$ is CA-S-observable with respect to $L(G)$, $\Sigma _{o}$, $\Sigma^a _{o}$, and $\Phi ^\pi$. Furthermore, if SCPDES-RL-CA is solvable, then $\cS_p$ defined in Equation (\ref{MB}) is a solution, that is, $L_r(\cS_p^a/G) = K$.
\end{theorem}

\noindent {\em Proof:} 

Note that
\begin{equation} \label{allgamma}
	\begin{split}
		&(\forall \gamma \in  \cS^a(t))\sigma \in \gamma  \\
		\Leftrightarrow
		&(\forall \gamma \in \Delta (\cS(t)))\sigma \in \gamma \\
		\Leftrightarrow
		& (\forall \gamma', \gamma'' \subseteq \Sigma^a_c) \sigma \in((\cS(t)-\gamma')\cup \gamma'') \\
		\Leftrightarrow
		&\sigma \in \cS(t) - \Sigma^a_c \\
		&(\mbox{since } \gamma' = \Sigma^a_c \wedge \gamma'' = \emptyset \mbox{ covers all cases}). 
	\end{split}
\end{equation}


Therefore,
\begin{align}
	\label{MBcc}
	&	s \sigma \in L_r(\cS^a/G)\nonumber\\
	\Leftrightarrow 
	& s\in L_r(\cS^a/G)\wedge s \sigma \in L(G) \wedge (\sigma \in \Sigma_{uc}  \nonumber\\
	&\vee  (\forall t \in   \Phi ^\pi(s))(\forall \gamma \in  \cS^a(t))\sigma \in \gamma)\nonumber\\
	\Leftrightarrow 
	&  s\in L_r(\cS^a/G)\wedge s \sigma \in L(G) \wedge (\sigma \in \Sigma _{uc}\nonumber \\
	&\vee (\forall t \in   \Phi ^\pi(s)) \sigma \in \cS(t) - \Sigma^a_c) \nonumber \\
	&(\mbox{by Equation (\ref{allgamma})}) \nonumber\\
	\Leftrightarrow 
	& s\in L_r(\cS^a/G)\wedge s \sigma \in L(G) \wedge (\sigma \in \Sigma _{uc}\nonumber\\
	&\vee (\forall t \in   \Phi ^\pi(s)) (\sigma \in \cS(t) \wedge \sigma \notin \Sigma^a_c)) \\
	\Leftrightarrow 
	& s\in L_r(\cS^a/G)\wedge s \sigma \in L(G) \wedge (\sigma \in \Sigma _{uc}\nonumber\\
	&\vee (\sigma \notin \Sigma^a_c \wedge (\forall t \in   \Phi ^\pi(s)) \sigma \in \cS(t) ))\nonumber \\
	\Leftrightarrow 
	& s\in L_r(\cS^a/G)\wedge s \sigma \in L(G) \wedge ((\sigma \in \Sigma _{uc} \vee \sigma \notin \Sigma^a_c)\nonumber\\
	& \wedge (\sigma \in \Sigma _{uc} \vee (\forall t \in   \Phi ^\pi(s)) \sigma \in \cS(t) )) \nonumber\\
	\Leftrightarrow 
	& s\in L_r(\cS^a/G)\wedge s \sigma \in L(G) \wedge \sigma \notin \Sigma^a_c \nonumber\\
	& \wedge (\sigma \in \Sigma _{uc} \vee (\forall t \in   \Phi ^\pi(s)) \sigma \in \cS(t) )\nonumber \\
	& (\mbox{because } \sigma \in \Sigma _{uc} \Rightarrow \sigma \notin \Sigma^a_c ).	\nonumber
\end{align}

We can now prove the theorem as follows.

(IF) Assume that $K$ is CA-S-controllable with respect to $L(G)$, $\Sigma _{uc}$, and $\Sigma^a _{c}$ and  CA-S-observable with respect to $L(G)$, $\Sigma _{o}$, and $\Sigma^a _{o}$, and $\Phi ^\pi$. We show that $\cS_p$ is a supervisor such that $L_r(\cS^a_p/G)=K$, that is, we prove, for all $s \in \Sigma^*$.
$$
s \in L_r(\cS^a_p/G) \Leftrightarrow s \in K
$$ 
by induction on the length $|s|$ of $s$.

{\em Base:} Since $K$ is nonempty and closed, $\varepsilon \in K \cap L(G)$. By definition, $\varepsilon \in L_r(\cS^a_p/G)$. Therefore, for $|s| = 0$, that is, $s=\varepsilon$, we have
$$
s \in L_r(\cS^a_p/G) \Leftrightarrow  s \in K.
$$

{\em Induction Hypothesis:} Assume that for all $s \in \Sigma^*$, $|s| \leq m$,
$$
s \in L_r(\cS^a_p/G) \Leftrightarrow s \in K.
$$

{\em Induction Step:} We show that for all $s \in \Sigma^*$, $\sigma \in \Sigma$, $|s\sigma| = m+1$,
$$
s \sigma \in L_r(\cS^a_p/G) \Leftrightarrow s \sigma \in K
$$
as follows.
\begin{align*}
	& s \sigma \in L_r(\cS^a_p/G) \\
	\Leftrightarrow 
	& s\in L_r(\cS^a_p/G)\wedge s \sigma \in L(G) \wedge \sigma \notin \Sigma^a_c \\
	& \wedge (\sigma \in \Sigma _{uc} \vee (\forall t \in   \Phi ^\pi(s)) \sigma \in \cS_p(t) ) \\
	& (\mbox{by Equation (\ref{MBcc})}) \\
	\Leftrightarrow 
	& \sigma \notin \Sigma^a_c \wedge s\in K \wedge s \sigma \in L(G) \\
	& \wedge (\sigma \in \Sigma _{uc} \vee (\forall t \in   \Phi ^\pi(s)) \sigma \in \cS_p(t) ) \\
	& (\mbox{by Induction Hypothesis}) \\
	\Leftrightarrow 
	& \sigma \notin \Sigma^a_c \wedge ((s\in K \wedge s \sigma \in L(G) \wedge \sigma \in \Sigma _{uc} )\\
	& \vee (s\in K \wedge s \sigma \in L(G) \wedge (\forall t \in   \Phi ^\pi(s)) \sigma \in \cS_p(t) )) \\
	\Leftrightarrow 
	& \sigma \notin \Sigma^a_c \wedge (s \sigma \in K \\
	& \vee (s\in K \wedge s \sigma \in L(G) \wedge (\forall t \in  \Phi ^\pi(s)) \sigma \in \cS_p(t) )) \\
	& (\mbox{by CA-S-controllability of $K$})\\
	\Leftrightarrow 
	& \sigma \notin \Sigma^a_c \wedge (s \sigma \in K \vee (s\in K \wedge s \sigma \in L(G) \\
	& \wedge (\forall t \in  \Phi ^\pi(s)) (\exists q \in
	SE^\pi_H(t))  \delta_{H} (q,\sigma) \in Q_H )) \\
	&(\mbox{by the definition of } \cS_p (t)) \\
	\Leftrightarrow 
	& \sigma \notin \Sigma^a_c \wedge (s \sigma \in K \vee (s\in K \wedge s \sigma \in L(G) \\
	& \wedge (\forall t \in  \Phi ^\pi(s)) (\exists s' \in K) t \in \Phi ^\pi(s') \wedge s' \sigma\in K) ) \\
	&(\mbox{by the definition of } {SE}^\pi_H(t)) \\
	\Leftrightarrow 
	& \sigma \notin \Sigma^a_c \wedge (s \sigma \in K \vee s \sigma \in K ) \\
	& (\mbox{by Lemma 1}) \\
	\Leftrightarrow 
	& s \sigma \in K \\
	& (\mbox{by CA-S-controllability of $K$}) .
\end{align*}

(ONLY IF)  Assume that there exists a  supervisor $\cS$ such that $L_r(\cS^{a}/G) = K$. We want to prove that $K$ is CA-S-controllable with respect to $L(G)$, $\Sigma _{uc}$, 
and $\Sigma^a _{c}$ and  CA-S-observable with respect to $L(G)$, $\Sigma _{o}$, and $\Sigma^a _{o}$, and $\Phi ^\pi$.

We first prove that $K$ is CA-S-controllable with respect to $L(G)$, $\Sigma _{uc}$, and $\Sigma^a _{c}$ by contradiction. 
Suppose that $K$ is not CA-S-controllable and there exists a supervisor $\cS$ such that $L_r(\cS^{a}/G))=K$. Since $K$ is not CA-S-controllable means either $K\Sigma_{uc} \cap L(G) \not\subseteq K$ or $K \not\subseteq (\Sigma - \Sigma^a_c)^*$. Because a non-networked supervisor is a special case of a networked supervisor, $K\Sigma_{uc} \cap L(G)  \subseteq K$  is a necessary condition for the existence of a non-networked supervisor. Hence, $K \Sigma_{uc} \cap L(G) \not\subseteq K$ cannot be true. Therefore $K \not\subseteq (\Sigma - \Sigma_c^a)^*$ must be true, that is,
\begin{align*}
	& (\exists s \in \Sigma^*) (\exists \sigma \in \Sigma) s \sigma \in K \wedge s \in (\Sigma - \Sigma^a_c)^* \wedge s \sigma \not\in (\Sigma - \Sigma^a_c)^* \\
	\Rightarrow
	& (\exists s \in \Sigma^*) (\exists \sigma \in \Sigma) s \sigma \in K \wedge s \in (\Sigma - \Sigma^a_c)^* \wedge \sigma \not\in \Sigma - \Sigma^a_c \\
	\Rightarrow
	& (\exists s \in \Sigma^*) (\exists \sigma \in \Sigma) s \sigma \in K \wedge s \in (\Sigma - \Sigma^a_c)^* \wedge \sigma \in \Sigma^a_c \\
	\Rightarrow
	& (\exists s \in \Sigma^*) (\exists \sigma \in \Sigma) s \sigma \in K \wedge s \in K \wedge s \in (\Sigma - \Sigma^a_c)^* \wedge \sigma \in \Sigma^a_c \\
	\Rightarrow
	& (\exists s \in \Sigma^*) (\exists \sigma \in \Sigma)   s \sigma \in L_r(\cS^{a}/G)  \wedge s \in L_r(\cS^{a}/G) \wedge \sigma \in \Sigma^a_c \\
	& (\mbox{because } L_r(\cS^{a}/G) = K ) \\
	\Rightarrow 
	& (\exists s \in \Sigma^*) (\exists \sigma \in \Sigma) s \sigma \in L_r(\cS^{a}/G) \wedge s \sigma \not\in L_r(\cS^{a}/G) \\
	& (\mbox{by Equation (\ref{MBcc}), } \sigma \in \Sigma^a_c \Rightarrow s \sigma \not\in L_r(\cS^{a}/G) ) ,
\end{align*}
which is a contradiction.

Next, we prove that $K$ is CA-S-observable with respect to $L(G)$, $\Sigma _{o}$,  $\Sigma^a _{o}$, and $\Phi ^\pi$ by contradiction. 
Suppose $K$ is CA-S-controllable with respect to $L(G)$, $\Sigma_{c}$, and $\Sigma^a _{c}$ but not  CA-S-observable with respect to  $L(G)$, $\Sigma _{o}$,  $\Sigma^a _{o}$, and $\Phi ^\pi$. 
By Equation (\ref{NSO}), we have
\begin{equation} \label{notobs}
	\begin{split}
		& \mbox{$K$ is not CA-S-observable} \\
		\Leftrightarrow
		& \neg (\forall s \in K)(\forall \sigma \in \Sigma) ((s\sigma \in L(G) \wedge(\forall t \in \Phi ^\pi(s)) \\
		& (\exists s' \in K) t \in  \Phi ^\pi(s') \wedge s'\sigma \in K) \Rightarrow s\sigma \in K) \\ 
		\Leftrightarrow
		& (\exists s \in K)(\exists \sigma \in \Sigma) \neg ((s\sigma \in L(G) \wedge(\forall t \in \Phi ^\pi(s)) \\
		& (\exists s' \in K) t \in  \Phi ^\pi(s') \wedge s'\sigma \in K) \Rightarrow s\sigma \in K) \\ 
		\Leftrightarrow
		& (\exists s \in K)(\exists \sigma \in \Sigma) s\sigma \in L(G) \wedge(\forall t \in \Phi ^\pi(s)) \\
		& (\exists s' \in K) t \in  \Phi ^\pi(s') \wedge s'\sigma \in K \wedge s\sigma \not\in K \\
		\Leftrightarrow
		& (\exists s \in K)(\exists \sigma \in \Sigma) \sigma \not\in \Sigma^a_c \wedge s\sigma \in L(G) \wedge(\forall t \in \Phi ^\pi(s)) \\
		& (\exists s' \in K) t \in  \Phi ^\pi(s') \wedge s'\sigma \in K \wedge s\sigma \not\in K \\
		& (\mbox{by Equation (\ref{Defc}), } s'\sigma \in K \Rightarrow \sigma \not\in \Sigma^a_c ).
	\end{split}
\end{equation}
Consider two possible cases for $\cS$.

{\em Case 1}: $(\forall t \in \Phi ^\pi(s)) \sigma \in \cS(t)$. In this case, by the derivation above, 
\begin{align*}
	& (\exists s \in K)(\exists \sigma \in \Sigma) \sigma \not\in \Sigma^a_c \wedge s\sigma \in L(G) \\
	& (\forall t \in \Phi ^\pi(s)) \sigma \in \cS(t) \wedge s\sigma \not\in K \\
	\Rightarrow
	& (\exists s \in \Sigma ^*)(\exists \sigma \in \Sigma) s \in K \wedge \sigma \not\in \Sigma^a_c \wedge s\sigma \in L(G) \\
	& \wedge (\forall t \in \Phi ^\pi(s)) \sigma \in \cS(t) \wedge s\sigma \not\in K \\ 
	\Rightarrow 
	& (\exists s \in \Sigma ^*)(\exists \sigma \in \Sigma) s \in L_r(\cS^a/G) \wedge \sigma \not\in \Sigma^a_c \wedge s\sigma \in L(G) \\
	& \wedge (\forall t \in \Phi ^\pi(s)) \sigma \in \cS(t) \wedge s\sigma \not\in K \\ 
	& (\mbox{because } L_r(\cS^a/G)=K) \\
	\Rightarrow 
	& (\exists s \in \Sigma ^*)(\exists \sigma \in \Sigma) s\in L_r(\cS^a/G)\wedge s \sigma \in L(G) \wedge \sigma \notin \Sigma^a_c \\
	& \wedge (\sigma \in \Sigma _{uc} \vee (\forall t \in   \Phi ^\pi(s)) \sigma \in \cS(t) ) \wedge s\sigma \not\in K \\
	\Rightarrow
	& (\exists s \in \Sigma ^*)(\exists \sigma \in \Sigma) s \sigma \in L_r(\cS^a/G) \wedge s\sigma \not\in K \\
	& (\mbox{by Equation (\ref{MBcc})}) ,
\end{align*}
which contradicts the assumption that $L_r(\cS^a/G)=K$. 

{\em Case 2}: $(\exists t \in \Phi ^\pi(s)) \sigma \not\in \cS(t)$ (=$ \neg (\forall t \in \Phi ^\pi(s)) \sigma \in \cS(t)$). In this case, by replacing $t$ with $t'$ in the derivation above, we have
\begin{align*}
	& (\exists s \in K)(\exists \sigma \in \Sigma) \sigma \not\in \Sigma^a_c \wedge s\sigma \in L(G) \\
	& \wedge (\exists t \in \Phi ^\pi(s)) \sigma \not\in \cS(t) \\
	& \wedge (\forall t' \in \Phi ^\pi(s)) (\exists s' \in K) t' \in  \Phi ^\pi(s') \wedge s'\sigma \in K \wedge s\sigma \not\in K \\
	\Rightarrow
	& (\exists s \in K)(\exists \sigma \in \Sigma) \sigma \not\in \Sigma^a_c \wedge s\sigma \in L(G) \wedge \sigma \not\in \Sigma_{uc} \\
	& \wedge (\exists t \in \Phi ^\pi(s)) \sigma \not\in \cS(t) \\
	& \wedge (\forall t' \in \Phi ^\pi(s)) (\exists s' \in K) t' \in  \Phi ^\pi(s') \wedge s'\sigma \in K \wedge s\sigma \not\in K \\
	& (\mbox{by Equation (\ref{Defc})}) \\
	& s \in K \wedge s\sigma \in L(G) \wedge s\sigma \not\in K \Rightarrow \sigma \not\in \Sigma_{uc} ) \\
	\Rightarrow
	& (\exists s \in \Sigma ^*)(\exists \sigma \in \Sigma) s \in K \wedge \sigma \not\in \Sigma^a_c \wedge s\sigma \in L(G) \\
	& \wedge (\exists t \in \Phi ^\pi(s)) \sigma \not\in \cS(t) \wedge \sigma \not\in \Sigma_{uc} \\
	& \wedge (\exists s' \in K) t \in  \Phi ^\pi(s') \wedge s'\sigma \in K \wedge s\sigma \not\in K \\
	& (\mbox{let } t'=t) \\
	\Rightarrow
	& (\exists s' \in \Sigma ^*) (\exists \sigma \in \Sigma) s' \in K \wedge \sigma \not\in \Sigma_{uc} \\
	& \wedge (\exists t \in \Phi ^\pi(s')) \sigma \not\in \cS(t) \wedge s'\sigma \in K \\
	\Rightarrow
	& (\exists s' \in \Sigma ^*) (\exists \sigma \in \Sigma) s' \in L_r(\cS^a/G) \wedge \sigma \not\in \Sigma_{uc} \\
	& \wedge (\exists t \in \Phi ^\pi(s')) \sigma \not\in \cS(t) \wedge s'\sigma \in K \\
	& (\mbox{because } L_r(\cS^a/G)=K) \\
	\Rightarrow
	& (\exists s' \in \Sigma ^*) (\exists \sigma \in \Sigma) s' \in L_r(\cS^a/G) \wedge \neg (\sigma \in \Sigma_{uc} \\
	& \vee (\forall t \in \Phi ^\pi(s')) \sigma \in \cS(t) ) \wedge s'\sigma \in K \\
	\Rightarrow
	& (\exists s' \in \Sigma ^*) (\exists \sigma \in \Sigma) s' \sigma \not\in L_r(\cS^a/G) \wedge s'\sigma \in K \\
	& (\mbox{by Equation (\ref{MBcc})}),
\end{align*}
which contradicts the assumption that $L_r(\cS^a/G)=K$. 

\hfill \bull

Let us illustrate the results using the following example.

\begin{example} 
	
Let us again consider the DES $G$ shown in Fig.~\ref{fig1}. The specification language  $K=L(H)$, where $H$ is the  sub-automaton of $G$ shown in Fig.~\ref{fig3}.
Assume that $\Sigma_o = \{\alpha, \beta, \mu,\lambda\}$ and $\Sigma^a_o = \{\alpha\}$.


The automata $F_{tr}$ marking language $A_{tr}$ for $tr$ $=$ $(3, \alpha, 4)$ is shown in Fig.~\ref{fig4}. Replace transition $tr$ with the corresponding automata $F_{tr}$ to obtain $H^{\diamond}$, which is shown in Fig.~\ref{fig5}. 
Automaton $H^\diamond_{\varepsilon}$ is constructed and shown in Fig.~\ref{fig6}. The CA-observer $H^\diamond_{obs}$ for $H$ is shown in Fig.~\ref{fig7}.

\begin{figure}[hbt!]
	\centering
	\includegraphics[height=0.8 in]{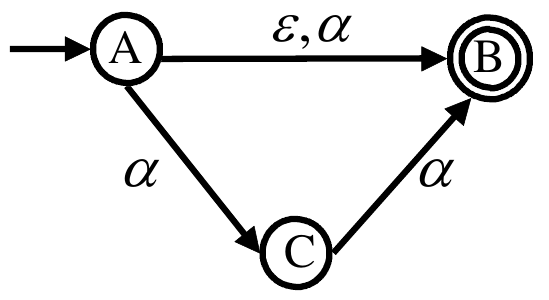}
	\caption{Automaton $F_{tr}$ marking language $A_{tr}$.}
	\label{fig4}
\end{figure}

\begin{figure}[hbt!]
	\centering
	\includegraphics[height=1.7 in]{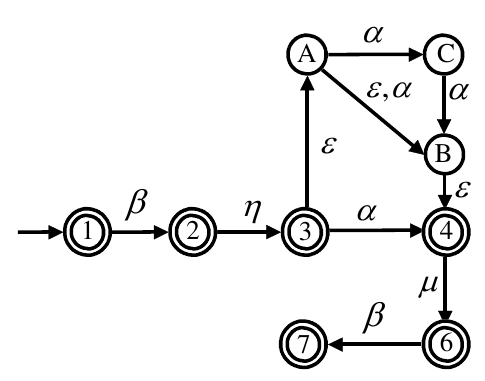}
	\caption{Automaton $H^\diamond$.}
	\label{fig5}
\end{figure}

\begin{figure}[hbt!]
	\centering
	\includegraphics[height=1.7 in]{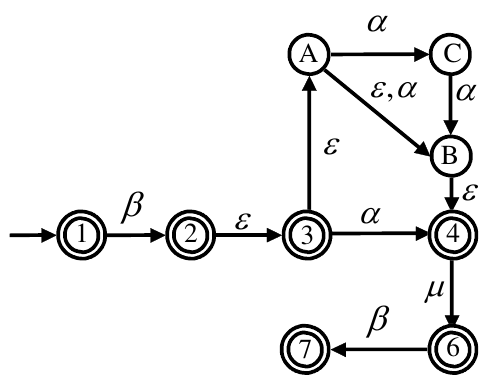}
	\caption{Automaton $H^\diamond_{\varepsilon}$ owing to $\Sigma_{uo} = \{\eta\}$.}
	\label{fig6}
\end{figure}

\begin{figure}[hbt!]
	\centering
	\includegraphics[height=1.1 in]{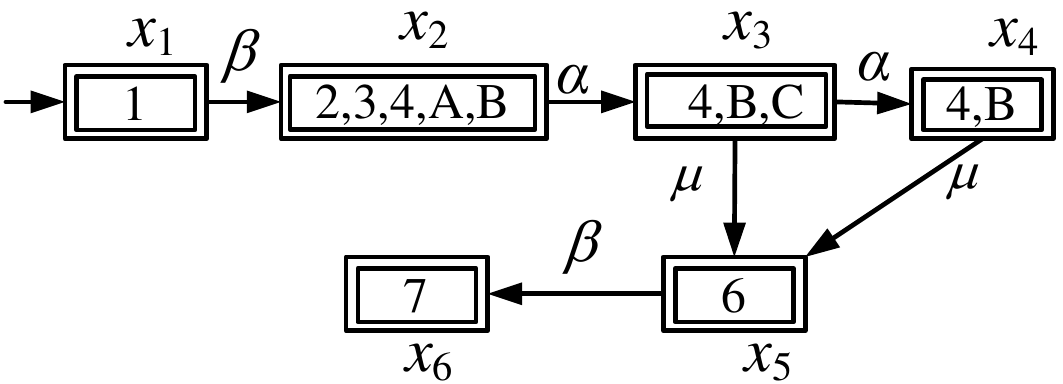}
	\caption{The CA-observer  $H^\diamond_{obs}$ for $H$.}
	\label{fig7}
\end{figure}

Based on the CA-observer  $H^\diamond_{obs}$, the state estimate for any observation $t\in \Phi ^\pi(L(H))$ can be calculated. For example, for $t = \beta\alpha$, we have 
$$
\xi_H(t) = \{4,B,C\}.
$$
Hence, ${SE}^\pi_H(t)$ $=$ $\xi_H (x_0, t) \cap Q_H$ $=$  $\{4,B,C\}$ $\cap Q_H$ $=$ $\{4\}$.

Consider the following two cases.

\noindent {\em Case 1}: Assume that $\Sigma_c= \Sigma$ and $\Sigma^a_c$ $=$ $\{\beta\}$. In this case, 
 ($\Sigma - \Sigma^a_{c})^* = \{\eta, \alpha, \mu, \lambda\}^*$, hence $K \not\subseteq (\Sigma - \Sigma^a_{c})^*$.

Therefore, $K$ is not CA-S-controllable with respect to $L(G)$, $\Sigma _{uc}$, and $\Sigma_c^a$.
By Theorem \ref{theo2},  no supervisor exists.

\noindent {\em Case 2}: Assume that  $\Sigma_c=\Sigma-\{\beta\}$ and $\Sigma^a_c$ $=$ $\emptyset$. In this case,  it can be checked Equation (\ref{Defc}) is satisfied. Hence, $K$ is CA-S-controllable with respect to $L(G)$, $\Sigma _{uc}$, and $\Sigma_c^a$.

\begin{table}[t]
	\centering
	\caption{Illustration of supervisor $\cS_{p}$}
	\begin{tabular}{|c|c|c|c|}
		\hline
		$t$  &   $\xi_H(x_0, t)$   & ${SE}^\pi_H(t)$   &    $\cS_p (t)$     \\
		\hline \hline 
		$\varepsilon $   &$x_1$  & $\{1\}$&  $\{\beta\}$ \\
		\hline
		$\beta$   &  $x_2$ &  $\{2,3,4\}$ &  $\{\alpha, \eta,\mu\}$\\
		\hline
		$\beta\alpha$   & $x_3$ & $\{4\}$ &$\{\mu\}$ \\
		\hline
		$\beta\alpha\mu$    & $x_5$ &$\{6\}$ &$\{\beta\}$ \\
		\hline
		$\beta\alpha\alpha$   & $x_4$ &$\{4\}$ &$\{\mu\}$ \\
		\hline
		$\beta\alpha\alpha\mu$  & $x_5$ &$\{6\}$ &$\{\beta\}$ \\
		\hline
		$\beta\alpha\mu\beta$ & $x_6$ &$\{7\}$ &$\emptyset$ \\
		\hline
		$\beta\alpha\alpha\mu\beta$       & $x_6$ &$\{7\}$ &$\emptyset$ \\
		\hline
	\end{tabular}
	\label{Tab_1}
	\begin{align*}
		& t - \mbox{string observed by supervisor}. \\
		& \xi_H(x_0, t) - \mbox{corresponding state in } H^\diamond_{obs}. \\
		&  {SE}^\pi_H(t) - \mbox{state estimate after observing } t. \\
		& \cS_p (t) - \mbox{control after observing } t.
	\end{align*}
	
\end{table}
 By Equation (\ref{NSO}), it can be checked that $K$ is CA-S-observable. Based on  Theorem \ref{theo2},  the supervisor $\cS^a_p$ exists such that $L_r(\cS^a_p/G) = K$. 
 The supervisor $\cS_{p}$ can be obtained from Fig.~\ref{fig7} and Equation~(\ref{MB}). The control given by $\cS_{p}$ is illustrated in Table~\ref{Tab_1}. 

\end{example}

\section{Infimal CA-S-$Controllable$ and CA-S-$Observable$ Superlanguage}

If the required language $K_r$ is CA-S-controllable and CA-S-observable, then we can design a supervisor $\cS$ such that $L_r(\cS^a/G) = K_r$. This is ideal. 
However, in practical systems, $K_r$ may not be CA-S-controllable and$/$or CA-S-observable. If it is not, we want to design a supervisor $\cS$ such that $L_r(\cS^a/G) \supseteq K_r$. 
Hence, we need to find a  superlanguage $M \supseteq K_r$ such that $M$ is CA-S-controllable and CA-S-observable. Clearly, there may exist more than one such $M$. 
To best approximate $K_r$, we want $M$ to be as small as possible. Therefore, we want to find the infimal CA-S-controllable and CA-S-observable superlanguage of $K_r$, whose existence is investigated below.
%

The following theorem show that CA-S-observability is preserved under intersection. 

\begin{theorem} 
	\label{theoremo} 
	\rm
	Let $K_i, i=1,2,...$, be CA-S-observable with respect to $L(G)$, $\Sigma _{o}$,  $\Sigma^a _{o}$, and $\Phi ^\pi$, then $\cap _{i} K_i$ is also  CA-S-observable with respect to $L(G)$, $\Sigma _{o}$,  $\Sigma^a _{o}$, and $\Phi ^\pi$.
\end{theorem}
\noindent {\em Proof:} 

Equivalently, let us prove that if $\cap _{i} K_i$ is not  CA-S-observable, then there exists $K_i$ such that $K_i$ is not CA-S-observable. Indeed, by Equation (\ref{notobs}),
\begin{align*}
& \mbox{$\cap _{i} K_i$ is not  CA-S-observable} \\
\Leftrightarrow
& (\exists s \in \cap _{i} K_i)(\exists \sigma \in \Sigma) (s\sigma \in L(G) \\
& \wedge(\forall t \in \Phi ^\pi(s)) (\exists s' \in \cap _{i} K_i) \\
&  t \in  \Phi ^\pi(s') \wedge s'\sigma \in \cap _{i} K_i) \wedge s\sigma \not\in \cap _{i} K_i \\
\Rightarrow
& (\exists j) (\exists s \in \cap _{i} K_i)(\exists \sigma \in \Sigma) (s\sigma \in L(G) \\
& \wedge(\forall t \in \Phi ^\pi(s)) (\exists s' \in \cap _{i} K_i) \\
&  t \in  \Phi ^\pi(s') \wedge s'\sigma \in \cap _{i} K_i) \wedge s\sigma \not\in K_j \\
& (\mbox{because } s\sigma \not\in \cap _{i} K_i \Rightarrow (\exists j) s\sigma \not\in K_j) \\
\Rightarrow
& (\exists j) (\exists s \in K_j) (\exists \sigma \in \Sigma) (s\sigma \in L(G) \\
& \wedge(\forall t \in \Phi ^\pi(s)) (\exists s' \in K_j) \\
&  t \in  \Phi ^\pi(s') \wedge s'\sigma \in K_i) \wedge s\sigma \not\in K_j \\
\Leftrightarrow
& \mbox{$(\exists j) K_j$ is not  CA-S-observable.}
\end{align*} 

\hfill \bull

We also show that CA-S-controllability is preserved under intersection as follows.

\begin{theorem}
	\label{Theorem4}
	\rm
	Let $K_i, i=1,2,...$, be CA-S-controllable with respect to $L(G)$, $\Sigma_{uc}$, and $\Sigma^a_c$, then $\cap _{i} K_i$ is also CA-S-controllable with respect to $L(G)$,  $\Sigma_{uc}$, and $\Sigma^a_c$.
\end{theorem}
\noindent {\em Proof:}

We need to prove
\begin{align*}
	& (\forall i) K_i \Sigma_{uc} \cap L(G) \subseteq K_i \wedge K_i \subseteq (\Sigma - \Sigma^a_c)^* \\
	\Rightarrow 
	& (\cap _{i} K_i) \Sigma_{uc} \cap L(G) \subseteq (\cap _{i} K_i) \wedge (\cap _{i} K_i) \subseteq (\Sigma - \Sigma^a_c)^* .
\end{align*}
Indeed, we have
\begin{align*}
	& (\cap _{i} K_i) \Sigma_{uc} \cap L(G) \\
	= 
	& (\cap _{i} K_i \Sigma_{uc}) \cap L(G) \\
	= 
	& \cap _{i} (K_i \Sigma_{uc} \cap L(G)) \\
	\subseteq & (\cap _{i} K_i) .
\end{align*}
Furthermore,
\begin{align*}
	& (\forall i) K_i \subseteq (\Sigma - \Sigma^a_c)^* \\
	\Rightarrow 
	& (\cap _{i} K_i) \subseteq (\Sigma - \Sigma^a_c)^* .
\end{align*}
\hfill \bull

	
	
	
	

In conventional supervisory control without cyber attacks, if $K_r \subseteq L(G)$ is not controllable and observable, then we can always find the (unique) infimal controllable and observable superlanguage of $K_r$. 
This is true because $L(G)$ is always controllable and observable. So, in the worst case, the infimal controllable and observable superlanguage of $K_r$ equals to $L(G)$.
	
The same, however, is not true for supervisory control under cyber attacks. This is because $L(G)$ may not be CA-S-controllable. To see this, consider the (least restrictive) supervisor $\cS_{lr}$ that enables all events. We have the following lemma.
\begin{lemma}
	\label{Lemma6}
	\rm
	The small language of the supervisor $\cS_{lr}(t) = \Sigma$, for all $t \in \Phi ^\pi(L(G))$, is given by 
	\begin{align*} 
		L_r(\cS_{lr}^a/G) = L(G) \cap (\Sigma - \Sigma^a_c)^*
	\end{align*} 
\end{lemma}
\noindent {\em Proof:} 

We prove that, for all $s \in \Sigma^*$,
\begin{align*} 
	s \in L_r(\cS_{lr}^a/G) \Leftrightarrow s \in L(G) \cap (\Sigma - \Sigma^a_c)^*
\end{align*} 
by induction on the length $|s|$ of $s$.

{\em Base:} Since $L(G)$ is nonempty and closed, $\varepsilon \in L(G) \cap (\Sigma - \Sigma^a_c)^*$. By definition, $\epsilon \in L_r(\cS_{lr}^a/G)$. Therefore, for $|s| = 0$, that is, $s=\varepsilon$, we have
$$
s \in L_r(\cS_{lr}^a/G) \Leftrightarrow s \in L(G) \cap (\Sigma - \Sigma^a_c)^*
$$

{\em Induction Hypothesis:} Assume that for all $s \in \Sigma^*$, $|s| \leq m$,
$$
s \in L_r(\cS_{lr}^a/G) \Leftrightarrow s \in L(G) \cap (\Sigma - \Sigma^a_c)^*
$$

{\em Induction Step:} We show that for all $s \in \Sigma^*$, $\sigma \in \Sigma$, $|s\sigma| = m+1$,
$$
s \sigma \in L_r(\cS_{lr}^a/G) \Leftrightarrow s \sigma \in L(G) \cap (\Sigma - \Sigma^a_c)^*
$$
as follows. By Equation (\ref{MBcc}),
\begin{align*} 
	&	s \sigma \in L_r(\cS_{lr}^a/G)\\
	\Leftrightarrow 
	& s\in L_r(\cS_{lr}^a/G)\wedge s \sigma \in L(G) \wedge \sigma \notin \Sigma^a_c \\
	& \wedge (\sigma \in \Sigma _{uc} \vee (\forall t \in   \Phi ^\pi(s)) \sigma \in \cS_{lr}(t) ) \\
	\Leftrightarrow 
	& s\in L_r(\cS_{lr}^a/G)\wedge s \sigma \in L(G) \wedge \sigma \notin \Sigma^a_c \\
	& (\mbox{because } \cS_{lr}(t) = \Sigma ) \\
	\Leftrightarrow 
	& s\in L(G) \cap (\Sigma - \Sigma^a_c)^* \wedge s \sigma \in L(G) \wedge \sigma \notin \Sigma^a_c \\
	& (\mbox{by Induction Hypothesis} ) \\
	\Leftrightarrow 
	& s \sigma \in L(G) \cap (\Sigma - \Sigma^a_c)^* .
\end{align*} 

\hfill \bull

To overcome the difficulty that $L(G)$ may not be CA-S-controllable, let
\begin{align}
	\label{na}
	& L_{na}(G)= L(G) \cap (\Sigma - \Sigma^a_c)^*.
\end{align} 
Then, by Theorem \ref{theo2} and Lemma \ref{Lemma6}, $L_{na}(G)$ is CA-S-controllable with respect to $L(G)$, $\Sigma _{uc}$, and $\Sigma^a _{c}$; 
and CA-S-observable with respect to $L(G)$, $\Sigma _{o}$, $\Sigma^a _{o}$, and $\Phi ^\pi$. Furthermore, $L_{na}(G)$ is the largest small language possible, that is, for any supervisor $\cS$,
$$
L_r(\cS^a/G) \subseteq L_{na}(G).
$$
Therefore, in the rest of the paper, we assume that the required language $K_r \subseteq L_{na}(G)$. 

Since both CA-S-controllability and CA-S-observability are preserved under intersection, the (unique) infimal CA-S-controllable and CA-S-observable superlanguage of $K_r \subseteq L_{na}(G)$ exists.
Formally, define the set of CA-S-controllable and CA-S-observable  superlanguages of $K_r \subseteq L_{na}(G)$ as
\begin{align*}
	\mbox{CACO} ({K_r}) = & \{{M} \subseteq {L_{na}(G)}: {K_r} \subseteq  \emph{M}  \ and \  \emph{M} \mbox{ is  closed,} \\
	&\mbox{CA-S-controllable with respect to $L(G)$,} \\
	&\mbox{$\Sigma _{uc}$, and $\Sigma_c^a$, and CA-S-observable } \\
	&\mbox{with respect to $L(G)$,  $\Sigma_{o}$,  and }\Phi ^\pi\}.  
\end{align*}

\begin{theorem} 
	\label{Theo5}
	\rm
	Let $K_r \subseteq L_{na}(G)$. The infimal element of CACO($K_r$), called the infimal CA-S-controllable and CA-S-observable superlanguage of $K_r$ and denoted by inf CACO($K_r$), exists and is given by
	$$
	\mbox{inf CACO}(K_r) = \underset{\emph{M} \in \rm CACO(K_r)}{\cap} \emph{M}.
	$$
\end{theorem}

\noindent {\em Proof:} 

The result follows from Theorems \ref{theoremo}, \ref{Theorem4}, and Lemma \ref{Lemma6}.

\hfill \bull



	
	
	


\section{Conclusion}


This paper investigates small languages in supervisory control of DES under cyber attacks. The main contributions of the paper are summarized as follows.
(1) Two new concepts, namely  CA-S-observability and CA-S-controllability, are introduced. 
(2) A necessary and sufficient condition for a supervisor to exist under cyber attacks whose small language is equal to a given required language is derived and proved. 
(3) It is proved that CA-S-observability and CA-S-controllability are preserved under intersection.
(4) The infimal CA-S-controllable and CA-S-observable superlanguage of a required language is shown to exist.

In the future, we will consider the range problem in supervisory control of discrete event systems under cyber attacks. We will investigate how to design a supervisor so that the supervised system is safe using the large language and can perform some basic tasks using the small language. 

\bibliographystyle{ieeetr}

\bibliography{Abstract}

\begin{biography}[{\includegraphics[width=1.0 in,height=1.25in, keepaspectratio]{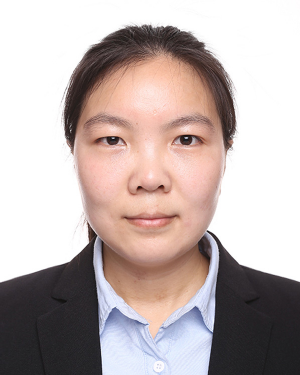}}]
	{Xiaojun Wang} received the B.Eng. degree in Electrical Engineering and Automation from Henan University of Urban Construction, China, in 2013, and the M.S. degree in Control Engineering from Kunming University of Science and Technology, China, in 2016, and Ph.D. degrees in Control Theory and Control Engineering from Xidian University, China, in 2021. She is currently a lecturer at University of Shanghai for Science and Technology.  Her research interests are discrete event systems (DES), networked DES, and their supervisory control techniques.
\end{biography}

\begin{biography}[{\includegraphics[width=1.0 in,height=1.25in, keepaspectratio]{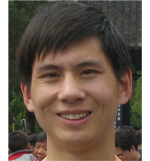}}]
	{Shaolong Shu} (M'12-SM'15) received his B.Eng. degree in automatic control, and his Ph.D. degree in control theory and control engineering from Tongji University,
	Shanghai, China, in 2003 and 2008, respectively. Since July, 2008, he has been with the School of Electronics and
	Information Engineering, Tongji University, Shanghai, China, where he is currently a full professor. From August, 2007
	to February, 2008 and from April, 2014 to April, 2015, he was a	visiting scholar in Wayne State University, Detroit, MI, USA.
	His main research interests include state estimation and control of discrete event systems and cyber-physical systems.
\end{biography}

\begin{biography}[{\includegraphics[width=1.0 in,height=1.25in, keepaspectratio]{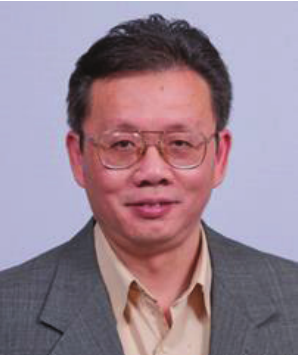}}]
	{Feng Lin} (S’85-M’88-SM’07-F’09) received his B.Eng. degree in electrical engineering from Shanghai Jiao Tong University, Shanghai, China, in 1982, and the M.A.Sc. and Ph.D. degrees in electrical engineering from the University of Toronto, Toronto, ON, Canada, in 1984 and 1988, respectively. He was a Post-Doctoral Fellow with Harvard University, Cambridge, MA, USA, from 1987 to 1988. Since 1988, he has been with the Department of Electrical and Computer Engineering, Wayne State University, Detroit, MI, USA, where he is currently a Professor. His current research interests include discrete event systems, hybrid systems, robust control, artificial intelligence, and their applications in alternative energy, biomedical systems, and automotive control. He authored a book entitled ``Robust Control Design: An Optimal Control Approach’’ and coauthored a paper that received a George Axelby outstanding paper award from the IEEE Control Systems Society. He was an associate editor of IEEE Transactions on Automatic Control.
\end{biography}

\end{document}